\pdfoutput=1
\documentclass[journal]{IEEEtran}
\usepackage{cite}
\usepackage{amsmath,amssymb,amsfonts}
\usepackage{algorithmic}
\usepackage{graphicx}
\usepackage{textcomp}
\usepackage{xcolor}
\usepackage[T1]{fontenc}
\usepackage{subcaption}
\usepackage{siunitx}
\usepackage{multirow}
\usepackage{float}
\usepackage{booktabs}
\usepackage{scalerel}
\usepackage{tikz}
\usetikzlibrary{svg.path}
\usepackage{threeparttable}
\definecolor{orcidlogocol}{HTML}{A6CE39}
\tikzset{
  orcidlogo/.pic={
    \fill[orcidlogocol] svg{M256,128c0,70.7-57.3,128-128,128C57.3,256,0,198.7,0,128C0,57.3,57.3,0,128,0C198.7,0,256,57.3,256,128z};
    \fill[white] svg{M86.3,186.2H70.9V79.1h15.4v48.4V186.2z}
                 svg{M108.9,79.1h41.6c39.6,0,57,28.3,57,53.6c0,27.5-21.5,53.6-56.8,53.6h-41.8V79.1z M124.3,172.4h24.5c34.9,0,42.9-26.5,42.9-39.7c0-21.5-13.7-39.7-43.7-39.7h-23.7V172.4z}
                 svg{M88.7,56.8c0,5.5-4.5,10.1-10.1,10.1c-5.6,0-10.1-4.6-10.1-10.1c0-5.6,4.5-10.1,10.1-10.1C84.2,46.7,88.7,51.3,88.7,56.8z};
  }
}

\newcommand\orcidicon[1]{\href{https://orcid.org/#1}{\mbox{\scalerel*{
\begin{tikzpicture}[yscale=-1,transform shape]
\pic{orcidlogo};
\end{tikzpicture}
}{|}}}}

\usepackage[hidelinks]{hyperref}

\hypersetup{
  colorlinks       = true,
  linkcolor        = {red!50!black},
  citecolor        = blue,
  urlcolor         = {blue!80!black}, 
  bookmarksnumbered= true,
  bookmarksopen    = true
}

\def\BibTeX{{\rm B\kern-.05em{\sc i\kern-.025em b}\kern-.08em
  T\kern-.1667em\lower.7ex\hbox{E}\kern-.125emX}}
\renewcommand{\arraystretch}{1.3}

\begin{document}

\title{Bit Error Rate and Performance Analysis of Multi-User OTFS under Nakagami–\(m\) Fading for 6G and Beyond Networks}

\author{%
  Emir Aslandogan\,\orcidicon{0000-0002-7158-7916}
  \and
   Haci Ilhan\,\orcidicon{0000-0002-6949-6126},\,
  \IEEEmembership{Senior Member,~IEEE}\\[2pt]
  \textit{Department of Electronics and Communication Engineering, Yildiz Technical University}\\
  Istanbul, Turkiye\\
  {\small\texttt{\{emira,ilhanh\}@yildiz.edu.tr}}%
  \thanks{This work was supported by the Scientific and Technological Research Council of Turkey (TÜBİTAK) under Project~123E513.}%
}
\maketitle
\begin{abstract}
Orthogonal Time-Frequency Space (OTFS) modulation stands out as a promising waveform for 6G and beyond wireless communication systems, offering superior performance over conventional methods, particularly in high-mobility scenarios and dispersive channel conditions. Error performance analysis remains crucial for accurately characterizing the reliability of wireless communication systems under practical constraints. In this paper, we systematically investigate the bit error rate (BER) performance of OTFS modulation over Nakagami–\(m\) fading channels in both single-user and multi-user scenarios. In analytical approaches, mathematical frameworks are employed for distinct receiver configurations: the Single-input Single-output (SISO) scenario leverages Erlang probability density function (PDF) of squared-Nakagami variables to derive closed-form BER expressions, while the Single-input Multiple-output (SIMO) case applies moment matching techniques with Gamma approximation to model multiple user interference, subsequently yielding Signal-to-interference-plus-noise Ratio (SINR) characterizations through Meijer–\(G\) functions. This study examines single-path and multi-path channel conditions, evaluating the relationship between path multiplicity and error performance metrics while considering various fading intensities through Nakagami–\(m\) fading parameters. The derived closed-form BER expressions are validated through maximum likelihood detection (MLD) based Monte Carlo simulations, demonstrating strong correlation between analytical and numerical results across various SNR regions. Furthermore, comparative benchmark evaluations against conventional orthogonal frequency division multiplexing (OFDM) with MLD reveal that OTFS consistently achieves superior error performance in high-mobility scenarios. In multipath fading environments, OTFS achieves superior diversity gain compared to conventional OFDM, which refers to enhanced error performance.
\end{abstract}


\section{Introduction}
\IEEEPARstart{O}{TFS} modulation has recently drawn considerable attention due to its substantial potential to deliver reliable communications in high-mobility scenarios, establishing itself as a highly promising scheme for wireless communication systems \cite{Orthogonal_Time_Frequency_Space_Modulation}, \cite{Li_2023}, \cite{New_Generation_of_Modulation_Addressing_the_Challenges}. Unlike traditional modulation techniques, OTFS applies a sequence of two-dimensional transformations to a doubly-dispersive channel, effectively converting it into an almost non-fading channel in the delay-Doppler domain \cite{Ramachandran_Chockalingam_2018, Li_2020}. High-mobility communications at elevated carrier frequencies are subject to considerable Doppler spread due to the relative motion among the transmitter, receiver, and surrounding scatterers. Conventional orthogonal frequency-division multiplexing (OFDM), widely employed in 4G, emerging 5G cellular systems, and WiFi networks, experiences performance degradation in such environments due to inter-carrier interference (ICI). This interference is further intensified by the notable disparity in normalized Doppler effects between the highest and lowest subcarriers, thereby complicating the synchronization process \cite{Orthogonal_Time-Frequency_Space_Modulation_A_Promising_Next_Generation_Waveform}. In contrast to the operation of OFDM within the time-frequency domain, OTFS employs a two-dimensional delay-Doppler domain to modulate information symbols, thereby exploiting the full potential of time-frequency diversity and conferring exceptional resilience to pronounced Doppler effects—a critical challenge in vehicular and satellite communication systems \cite{9518377}. 

Owing to these superior characteristics, OTFS modulation has emerged as a prominent area of research in wireless communications. The extensive body of research on OTFS addresses a wide array of critical areas, including fundamental signal processing challenges such as channel estimation \cite{CE1}, \cite{CE2}, \cite{CE3} and equalization techniques \cite{EQ1}, \cite{EQ2}, \cite{EQ3} the design of low-complexity receivers \cite{LCR1,LCR2}, and comprehensive error performance analyses \cite{ERR1}, \cite{ERR2}, \cite{ERR3}, often coupled with its integration with various channel coding schemes like Low-density parity-check code
(LDPC) \cite{LDPC1}, Polar \cite{POL1}, and Turbo codes \cite{TUR1}. Furthermore, studies examine robust pulse shaping \cite{PS1,PS2} and windowing methods \cite{WW1,WW2}, the application of OTFS in Multiple-Input Multiple-Output (MIMO) systems \cite{MIMO1,MIMO2}, resource allocation approaches, such as Non-Orthogonal Multiple Access (NOMA) integrated with OTFS (OTFS-NOMA) \cite{NOMA1,NOMA2}, the development of OTFS with index modulation (OTFS-IM) \cite{IM1}, \cite{IM2}, \cite{otfs_media} and considerations for security and privacy \cite{SEC1}. Considering this broad spectrum of research areas and the performance enhancements promised by OTFS, a detailed error performance analysis of this modulation technique, particularly under diverse and realistic channel conditions, is of critical importance.

In this paper, we make the following key contributions:
\begin{itemize}
    \item We systematically investigate the bit error rate (BER) performance of OTFS modulation over Nakagami–$m$ fading channels, a generalized fading model, in both single-user (SISO) and multi-user (SIMO) scenarios. Addressing a notable gap in detailed analytical BER studies for OTFS over various fading models, this work provides, to the best of our knowledge, the first comprehensive analytical and simulation-validated BER analysis specifically for OTFS over Nakagami-$m$ fading channels.
    \item For the (SISO) OTFS system, we derive novel closed-form BER expressions by leveraging the Erlang probability density function (PDF) of squared-Nakagami variables.
    \item For the multi-user interference scenario, modeled within a SIMO-OTFS framework, we apply moment matching techniques with Gamma approximation to characterize the Signal-to-Interference-plus-Noise Ratio (SINR) using Meijer–\(G\) functions, subsequently enabling BER performance evaluation.
    \item We conduct a comprehensive analysis of OTFS performance under both single-path and multi-path Nakagami-$m$ channel conditions, evaluating the impact of path multiplicity and varying fading intensities (Nakagami–\(m\) parameter) on error metrics.
    \item The accuracy of our derived analytical closed-form BER expressions is rigorously validated through extensive Monte Carlo simulations using maximum likelihood detection (MLD), demonstrating a strong correlation between analytical and numerical results across various Signal-to-Noise Ratio (SNR) regions.
    \item We provide comparative benchmark evaluations of OTFS against conventional OFDM with MLD, quantitatively demonstrating the consistent and superior error performance of OTFS in high-mobility Nakagami–\(m\) fading scenarios.
    \item We analyze the diversity gain performance of OTFS in SISO and SIMO scenarios and demonstrate its superiority over conventional OFDM in single path and multipath fading environments.
\end{itemize}
The remainder of this paper is organized as follows.
In Section~\ref{sec:background_otfs}, we provide a background on the OTFS modulation, detailing the transmitter architecture, the channel model, and the receiver structure.
Section~\ref{sec:ber_analysis} is dedicated to the BER analysis under Nakagami-$m$ fading channels, where we derive analytical expressions for the single-user case and subsequently for the multi-user case.
The performance of the proposed analytical frameworks is validated through extensive Monte Carlo simulations, and these numerical results are presented and discussed in Section~\ref{sec:numerical_results}, covering both SISO-OTFS and SIMO-OTFS performance.
Finally, Section~\ref{sec:conclusion} concludes the paper by summarizing our key findings and contributions.

\textit{Notations:} Scalars are designated by italic letters (e.g., $x, M, P, m_i$). Vectors and matrices are denoted using bold lowercase (e.g., $\mathbf{s}, \mathbf{x}, \mathbf{r}, \mathbf{w}$) and bold uppercase (e.g., $\mathbf{X}, \mathbf{H}, \mathbf{S}, \mathbf{G}_{tx}, \mathbf{F}_N, \mathbf{I}_M, \mathbf{\Pi}, \mathbf{\Delta}$) characters, respectively. The set of complex numbers is represented by $\mathbb{C}$. Matrix vectorization and its inverse operation are denoted by $\text{vec}(\cdot)$ and $\text{vec}^{-1}(\cdot)$, respectively. The $N \times N$ discrete Fourier transform (DFT) matrix and the $M \times M$ identity matrix are represented by $\mathbf{F}_N$ and $\mathbf{I}_M$, respectively. The Kronecker product is symbolized by $\otimes$. The Hermitian transpose operation is denoted by $(\cdot)^\dagger$. The statistical expectation operator is represented by $\mathbb{E}[\cdot]$. $\left[\cdot \right]_n$ denotes the modulo-$n$ operation The probability density function (PDF) and cumulative distribution function (CDF) of a random variable $X$ are designated $f_X(x)$ and $F_X(x)$, respectively. The Gamma function, lower incomplete Gamma function, and upper incomplete Gamma function are represented by $\Gamma(\cdot)$, $\gamma(\cdot,\cdot)$, and $\Gamma(\cdot,\cdot)$, respectively. The Meijer–\(G\) function is denoted by $G_{p,q}^{m,n}\left(\cdot \middle| \begin{smallmatrix} \cdot \\ \cdot \end{smallmatrix}\right)$. The double factorial is represented as $n!!$. The imaginary unit $\sqrt{-1}$ is denoted by $j$. The symbol $\triangleq$ signifies "is defined as".
\section{Background on OTFS Modulation}
\label{sec:background_otfs}
\subsection{Transmitter}
The multiplexing of \( MN \) information symbols operates on a delay-Doppler (DD) grid, where they are denoted as \( x[k,l] \). In this case, \( k = 0, \dots, N-1 \) represents the Doppler bins, whereas \( l = 0, \dots, M-1 \) indicates the delay bins. The parameters \( M \) and \( N \) denote the number of delay and Doppler bins, respectively. The time-frequency grid (TF grid) is represented as \( \Lambda \), with the subcarrier width defined as \( \Delta f = \frac{1}{T} \). The DD grid is defined as \( \Gamma = \left\{\left( \frac{k}{NT},\frac{l}{M \Delta f} \right) \mid k=0,\dots,N-1, \; l=0,\dots,M-1 \right\} \), where \( M\Delta f \) denotes the system bandwidth, and \( NT \) represents the OTFS frame duration \cite{8835764}. By applying the inverse symplectic finite Fourier transform (ISFFT), the information symbols in the DD domain are mapped onto the time-frequency (TF) grid and represented as \( X[n,m] \). Within the TF domain, these symbols are transformed into time-domain signals using the Heisenberg transform before being transmitted through the channel. At the receiver, the received time-domain signal is first converted back into TF domain symbols via the Wigner transform. Subsequently, a symplectic finite Fourier transform (SFFT) is employed to recover the original symbols in the DD domain. The TF domain symbols at the output of the ISFFT are given as \cite{8424569},
\begin{equation}
X\left [ n,m \right ] = \frac{1}{\sqrt{MN}}\sum_{k=0}^{N-1}\sum_{l=0}^{M-1}x\left [ k,l \right ]e^{j2\pi\left ( \frac{nk}{N}-\frac{ml}{M} \right )}.
\label{eq:1}
\end{equation}
The signal \( X\left[ n,m \right] \) is transformed into the time-domain signal using the Heisenberg transform,
\begin{equation}
s\left( t \right) = \sum_{n=0}^{N-1}\sum_{m=0}^{M-1}X\left[n,m\right]g_{\text{tx}}\left(t-nT  \right)e^{j2\pi\Delta f\left( t-nT \right)},
\label{eq:2}
\end{equation}
where $g_{\text{tx}}$ represents the transmitter pulse signal with a duration of \( T \). OTFS data transmission is performed by representing the signal \( s(t) \) as a discrete-time signal. In this process, samples of \( s(t) \) are taken at intervals of \( T/M \). Consequently, the discrete representation of \( s(t) \) is expressed as follows,
\begin{equation}
\mathbf{s} = \left\{ s\left( 0 \right), s\left( 1 \right), \cdots, s\left( MN-1 \right) \right\}.
\end{equation}
Furthermore, the modulated signal \( x[k,l] \) is denoted in matrix form and is defined as \( \mathbf{X} \in \mathbb{C}^{M \times N} \),
\begin{equation}
\mathbf{X} =
\begin{bmatrix}
x(0,0) & x(1,0) & \cdots & x(N{-}1,0) \\
x(0,1) & x(1,1) & \cdots & x(N{-}1,1) \\
\vdots & \vdots & \ddots & \vdots \\
x(0,M{-}1) & x(1,M{-}1) & \cdots & x(N{-}1,M{-}1)
\end{bmatrix}.
\label{eq:X_matrix}
\end{equation}

The transmitted symbol \( \mathbf{S} \in \mathbb{C}^{M \times N} \) can be represented as follows,
\begin{equation}
\mathbf{S} = \mathbf{G}_{\text{tx}}\mathbf{F}^{{\dagger}}_M\left( \mathbf{F}_M \mathbf{X} \mathbf{F}^{{\dagger}}_N\right)= \mathbf{G}_{\text{tx}}\mathbf{X} \mathbf{F}^{{\dagger}}_N,
\end{equation}
The pulse shaping waveform $g_{\text{tx}}$ is sampled at the sampling frequency $f_s = M/T$ to construct the diagonal matrix $\mathbf{G}_{\text{tx}}$, which is expressed as  
$\mathbf{G}_{\text{tx}} = \text{diag} \left[ g_{\text{tx}}(0), g_{\text{tx}}(T/M), \dots, g_{\text{tx}}((M-1)T/M) \right] \in \mathbb{C}^{M \times M}$. The function $g_{\text{tx}}(t)$ represents a rectangular pulse, which is defined as  
$g_{\text{tx}}(t) =  
\begin{cases} 
1, & \text{if }  0 \leq t \leq T, \\ 
0, & \text{otherwise}
\end{cases}.$ As a discrete-time signal, $\mathbf{G}_{\text{tx}}$ corresponds to the identity matrix $\mathbf{I}_M$ in the case of a rectangular pulse.
There exists the following relationship between the discrete-time signal $\mathbf{s}$ and the transmitted signal $\mathbf{S}$,
\begin{equation}
\mathbf{s} = \text{vec}\left( \mathbf{S} \right) = \left( \mathbf{F}^{\dagger}_N \otimes \mathbf{G}_{\text{tx}}  \right)\mathbf{x},
\end{equation}
where $\mathbf{x} = \text{vec}(\mathbf{X})$.
\subsection{Channel}
In a time-varying channel, the signal $s(t)$ propagates through the DD domain channel characterized by the response $h(\tau, \nu)$. Therefore, the received signal $r(t)$ is represented as,
\begin{equation}
r\left ( t \right ) = \int \int h\left ( \tau, \nu \right )s\left ( t-\tau \right )e^{j2\pi \nu \left ( t-\tau \right )}d \tau d \nu +  w\left ( t \right )
\end{equation}
where $w(t)$ represents the additive white Gaussian noise (AWGN). Considering a transmission channel with $P$ propagation paths, let $h_p$ denote the complex path gain of the $p$-th path, while $\tau_p$ and $\nu_p$ represent the corresponding delay and Doppler shifts, respectively. For $p \in \mathbb{N}[1, P]$, the delay-Doppler channel response $h(\tau, \nu)$ is expressed as,
\begin{equation}
h\left( \tau, \nu \right) = \sum_{p=1}^{P}h_p\delta\left( \tau-\tau_p \right)\delta\left( \nu-\nu_p \right),
\end{equation}
where $\tau_p$ and $\nu_p$ on the DD grid can be expressed as $\Gamma_p = \left( \tau_p,\nu_p \right) = \left( \frac{l_p}{M\Delta f},\frac{k_p}{NT} \right)$.
\subsection{Receiver}
The received signal $r(t)$ at the receiver is sampled at the frequency $f_s = M/T$, resulting in the discrete signal $\mathbf{r}$,
\begin{equation}
\mathbf{r} = \left\{ r\left( 0 \right), r\left( 1 \right), \cdots, r\left( MN-1 \right) \right\} =\left\{ r\left( n \right) \right\}^{MN-1}_{n=0}.
\end{equation}
The received signal $r(n)$ can be expressed in terms of the complex channel gain as  
\begin{equation}
r\left( n \right) = \sum_{p=1}^{P}h_pe^{j2\pi\frac{k_p\left( n-l_p \right)}{MN}}s\left( \left[ n-l_p \right]_{MN} \right)+w\left( n \right),
\label{eq:9}
\end{equation} 
The equation Eq. (\ref{eq:9}) can be represented in vectorized form as,
\begin{equation}
\mathbf{r} = \mathbf{H} \mathbf{x} + \mathbf{w},
\end{equation}
where $\mathbf{H}$ denotes the time-domain channel matrix. The channel matrix $\mathbf{H}$, which is a $MN \times MN$ matrix, can be defined as,
\begin{equation}
\mathbf{H} \triangleq \sum_{p=1}^{P}h_p \Pi^{l_p}\Delta^{k_p+\kappa_p}, 
\label{eq:11}
\end{equation}
where $\mathbf{\Pi}$ represents the permutation matrix, $\mathbf{\Delta}$ denotes the diagonal matrix is denoted, and $\kappa_p$ corresponds to the Doppler shift for the $p$-th path.
The matrix $\mathbf{\Pi}$ characterizes the delay effect and has the form as,
\begin{equation}
\mathbf{\Pi} =
\begin{bmatrix}
0 & \cdots & 0 & 1 \\
1 & \ddots & 0 & 0 \\
\vdots & \ddots & \ddots & 0 \\
0 & \cdots & 1 & 0
\end{bmatrix}_{MN \times MN}.
\end{equation}
Similarly, the $MN \times MN$ matrix $\mathbf{\Delta}$ is a diagonal matrix that characterizes the Doppler effect and consists of elements $\alpha = e^{\frac{j2\pi}{MN}}$, expressed as $\mathbf{\Delta} = \text{diag}\left\{\alpha^0, \alpha^1, \cdots , \alpha^{MN-1} \right\}$.
At the receiver, the received signal samples $\mathbf{r}$ are transformed into time-frequency domain symbols as $\mathbf{R}=\text{vec}^{-1}\left( \mathbf{r} \right)$. Here, the $\text{vec}^{-1}(\cdot)$ operation reshapes the $MN \times 1$ vector $\mathbf{r}$ into an $M \times N$ matrix $\mathbf{R}$. The receiver pulse shaping waveform $g_{\text{rx}}$ is sampled at the sampling frequency $f_s = M/T$ to construct the diagonal matrix $\mathbf{G}_{\text{rx}}$, which is expressed as  
$\mathbf{G}_{\text{rx}} = \text{diag} \left[ g_{\text{rx}}(0), g_{\text{rx}}(T/M), \dots, g_{\text{rx}}((M-1)T/M) \right] \in \mathbb{C}^{M \times M}$. 
The received DD domain signal at the receiver is represented as \cite{8516353}, 
\begin{align}
\mathbf{y} &= (\mathbf{F}_N \otimes \mathbf{G}_{\text{rx}}) \mathbf{r} \nonumber \\
&= (\mathbf{F}_N \otimes \mathbf{G}_{\text{rx}}) \mathbf{H} (\mathbf{F}_N^\dagger \otimes \mathbf{G}_{\text{tx}}) \mathbf{x}  
+ (\mathbf{F}_N \otimes \mathbf{G}_{\text{rx}}) \mathbf{w} \nonumber \\
&= \mathbf{H}_{\text{eff}} \mathbf{x} + \tilde{\mathbf{w}}.
\end{align}
where $\mathbf{H}_{\text{eff}} = (\mathbf{F}_N \otimes \mathbf{G}_{\text{rx}}) \mathbf{H} (\mathbf{F}_N^\dagger \otimes \mathbf{G}_{\text{tx}})$ represents the effective channel matrix, and $\tilde{\mathbf{w}} = (\mathbf{F}_N \otimes \mathbf{G}_{\text{rx}}) \mathbf{w}$ denotes the noise vector.
\section{Bit Error Rate Analysis Under Nakagami–\(m\) Fading Channel}
\label{sec:ber_analysis}
\subsection{Single User Case}
We are assuming $h_p$ values follow the Nakagami–\(m\) distribution, with $h_p$ denoting the channel coefficients with following PDF as \cite{Nakagami1960}, 
\begin{equation}
f_{\left|h_p \right|}\left ( z;m_i, \mu_i \right ) = \frac{2z^{2m_i-1}}{\mu_i^{m_i}\left ( m_i-1 \right )!}\text{exp}\left ( -\frac{z^2}{\mu_i} \right ), z\geq 0,
\end{equation}
where $m_i$ denotes Nakagami–\(m\) fading parameter, is a positive integer, and $\mu_i=E\left [ h_p^2 \right ]/m_i$. We first require the PDF of the square of the Nakagami–\(m\) distributed channel coefficients. $\left|h_p \right|^2$ follows the Erlang distribution as \cite{Kara_2006},
\begin{equation}
f_{\left|h_p \right|^2}\left ( z; m_i, \mu_i \right )=\frac{z^{m_i-1}}{\mu_i^{m_i}\left ( m_i-1 \right )!}\text{exp}\left ( -\frac{z}{\mu_i} \right ) , z\geq 0.
\label{eq:PDF_hp2}
\end{equation}
The CDF for the expression given in Eq. (\ref{eq:PDF_hp2}) can be specified as follows \cite{Alouini_2005},
\begin{equation}
F_{\left|h_p \right|^2}\left ( z; m_i, \mu_i \right )=  1-\frac{\Gamma\left ( m_i, \frac{z}{\mu_i} \right )}{\left ( m_i-1 \right )!} , z\geq 0,
\end{equation}
where $\Gamma\left ( \cdot , \cdot \right )$ is the incomplete Gamma function defined in \cite[eq. (8.350.2)]{Gradstejn_2000}. Using \cite[eq. (8.352.2)]{Gradstejn_2000}, CDF rewritten as,
\begin{equation}
F_{\left|h_p \right|^2}\left ( z; m_i, \mu_i \right ) = 1-\text{exp}\left ( -\frac{z}{\mu_i} \right )\sum_{i=1}^{m_i-1}\left ( \frac{1}{i!} \right )\left ( \frac{z}{\mu_i}\right )^i.
\label{eq:4}
\end{equation}
Since the signal arrives via $P$ paths, there are $P$ Nakagami–\(m\) channel coefficients, and the PDF function is obtained by summing the squares of these Nakagami–\(m\) channel coefficients as follows \cite{Kara_2006},
\begin{equation}
\begin{split}
f_{Z_P} &= \sum_{i=1}^{P}\sum_{k=1}^{m_i}\Xi\left( i, k, \left\{ m_q \right\}_{q=1}^{P}, \left\{ \mu_q \right\}_{q=1}^{P}, \left\{ p_q \right\}_{q=1}^{P-2} \right)\\
&\quad\times f_{\left| h_p \right|^2} \left( z; k, \mu_i \right),
\end{split}
\label{eq:5}
\end{equation}
where, $\Xi\left( i, k, \left\{ m_q \right\}_{q=1}^{P}, \left\{ \mu_q \right\}_{q=1}^{P}, \left\{ p_q \right\}_{q=1}^{P-2} \right)$ to be written as for simplicity $\Xi$ and $R_P \triangleq \sum_{i=1}^{P}m_i$. Also, $U(a) $ indicates the unit step function and $U(a>0) = 1$, otherwise equal to zero.  
\begin{equation}
\begin{split}
\Xi = &\sum_{i=1}^{P}\sum_{p_1=k}^{m_i}\sum_{p_2=k}^{p_1}\cdots \sum_{p_{P-2}=k}^{p_{P-3}}\\
&\times \left[ \frac{\left( -1 \right)^{R_{P}-m_i}\mu_i^k}{\prod_{h=1}^{P}\mu_h^{m_h}} \right.\\
&\times \frac{\left( m_i+m_{1+\text{U}\left( 1-i \right)}-p_1-1 \right)!}{\left( m_{1+\text{U}\left( 1-i \right)}-1 \right)!\left(m_i-p_1  \right)!}\\
&\times \left(  \frac{1}{\mu_i}-\frac{1}{\mu_{1+\text{U}\left( 1-i \right)}}\right)^{p_1-m_i-m_{1+\text{U}\left( 1-i \right)}} \\
&\times \frac{\left(p_{P-2}+m_{P-1+\text{U}\left( P-1-i \right)}-k-1  \right)!}{\left(m_{P-1+\text{U}\left( P-1-i \right)}-1  \right)!\left( p_{P-2} -k\right)!} \\
&\times \left( \frac{1}{\mu_i} -\frac{1}{\mu}_{P-1+\text{U}\left( P-1-i \right)}\right)^{k-p_{P-2}-m_{P-1+\text{U}\left( P-1-i \right)}}\\
&\times \prod_{d=1}^{P-3}\frac{\left(p_d+m_{d+1+\text{U}\left( d+1-i \right)}-p_{d+1} -1 \right)!}{\left(m_{d+1+\text{U}\left( d+1-i \right)}-1  \right)!\left(p_d - p_{d+1}  \right)!} \\
&\left.\times \left( \frac{1}{\mu_i} - \frac{1}{\mu_{d+1+\text{U}\left( d+1-i \right)}}\right)^{p_{d+1}-p_d -m_{d+1+\text{U}\left( d+1-i \right)}}\right].
\end{split}
\label{eq:21}
\end{equation}
Let $X_l$ for $l = 1,\ldots,L$ be independent Nakagami–\(m\) random variables with shape parameters $m_l$ and  $  \mu_{l} = \frac{\mathbb{E}[X_{l}^{2}]}{m_{l}}$. The squared magnitudes $Y_l = X_l^2$ follow Gamma distributions,
The sum of these squared magnitudes is given by:
\begin{equation}
Z_L = \sum_{l=1}^L Y_l
\end{equation}
where $m_i \neq m_j$ for all $i \neq j$. The probability density function (PDF) of $Z_L$ can be expressed as \cite{10794219},
\begin{equation}
\begin{split}
f_{Z_L}(z)
&= \sum_{i=1}^{L}\sum_{k=1}^{m_i}
\Xi_{L}\bigl(
i,\,k,\,
\{m_{q}\}_{q=1}^{L},\,
\{\eta_{q}\}_{q=1}^{L},\,
\{\iota_{q}\}_{q=1}^{L-2}
\bigr)\\
&\quad\times
f_{Y_i}\bigl(z;\,k,\,\mu_i\bigr).
\end{split}
\end{equation}
For an modulated signal in a system employing Maximal Ratio Combining (MRC), the average SER can be expressed as \cite{Tellambura_2004}, \cite{McKay_2007}, \cite{ALTUNBAS2012841},
\begin{equation}
\overline{P}_{e}^{\text{MRC}}
\;=\;
\frac{\mathcal{A}\,\sqrt{\mathcal{B}}}{2\,\sqrt{\pi}}
\int_{0}^{\infty}
y^{-\tfrac{1}{2}}\,
F_{\Upsilon}^{\text{MRC}}(y)\,
e^{-\,\mathcal{B}\,y}
\,\mathrm{d}y,
\label{eq:MRC_SER}
\end{equation}
where \(F_{\Upsilon}^{\text{MRC}}(y)\) denotes the CDF of the MRC‐combined SNR. As tabulated in Table \ref{Tablo}, the constants $\mathcal{A}$ and $\mathcal{B}$ are specified for the respective modulation schemes.
\begin{table}[ht]
\centering
\caption{Parameters $\mathcal{A}$ and $\mathcal{B}$ for $P_s(\gamma) \approx \mathcal{A}Q(\sqrt{2\mathcal{B}\gamma})$ for Several Signaling Constellations} 
\label{tab:ABparams_updated}
\begin{threeparttable}
\begin{tabular*}{\columnwidth}{@{\extracolsep{\fill}} l l l l @{}} 
\toprule
Modulation Scheme        & $\mathcal{A}$               & $\mathcal{B}$                 & $\mathcal{M}$  \\ 
\midrule
BPSK                     & $1$                         & $1$                           & ---                   \\
BFSK (coherent)          & $1$                         & $0.5$                         & ---                   \\
GMSK (approx. as MSK)    & $1$                         & $1$                           & ---                   \\
M-DEPSK (coherent)\tnote{*} & $2$                         & $\sin^2(\pi/\mathcal{M})$               & $\ge 2$               \\
QPSK                     & $2$                         & $0.5$                         & $\mathcal{M}=4$                 \\
M-PSK (approx. coherent)\tnote{*} & $2$                         & $\sin^2(\pi/\mathcal{M})$               & $\ge 2$               \\
M-FSK (coherent orth.)   & $\mathcal{M}-1$                       & $0.5$                         & $> 2$                 \\
Square M-QAM (approx.)   & $4(1-1/\sqrt{\mathcal{M}})$           & $1.5/(\mathcal{M}-1)$                   & $\ge 4$               \\
M-DPSK (non-coh. approx.)& $2$                         & $\sin^2(\pi/2\mathcal{M})$              & $\ge 2$               \\
DBPSK (non-coh. approx.) & $2$                         & $0.5$                         & $\mathcal{M}=2$                 \\
M-PAM                    & $2(\mathcal{M}-1)/\mathcal{M}$                  & $3/(\mathcal{M}^2-1)$                   & $\ge 2$               \\
\bottomrule
\end{tabular*}
\begin{tablenotes}
    \item[*] For M-PSK and M-DEPSK, these are common approximations for Symbol Error Rate (SER). BPSK is M-PSK with $M=2$ but typically uses $\mathcal{A}=1$. QPSK is M-PSK with $M=4$.
\end{tablenotes}
\end{threeparttable}
\label{Tablo}
\end{table}
The signalling formats investigated in this work, for which the parameters $\mathcal{A}$ and $\mathcal{B}$ (used in the context of the symbol error rate (SER) approximation $P_s(\Upsilon) \approx \mathcal{A}Q(\sqrt{2\mathcal{B}\Upsilon})$) are also tabulated in Table~\ref{tab:ABparams_updated}\tnote{*}, comprise: %
binary phase-shift keying (BPSK); %
binary frequency-shift keying (BFSK) with coherent detection; %
Gaussian minimum-shift keying (GMSK), with performance approximated by that of minimum shift keying (MSK); %
\(M\)-ary differentially encoded phase-shift keying (M-DEPSK) with coherent detection, for which the SER is approximated similarly to M-PSK; %
quadrature phase-shift keying (QPSK), treated as a special case of M-PSK (\(M=4\)); %
\(M\)-ary phase-shift keying (M-PSK) with coherent detection, employing established SER approximations; %
\(M\)-ary frequency-shift keying (M-FSK) with coherent orthogonal signaling, based on union bound approximations for SER; %
square \(M\)-ary quadrature amplitude modulation (square M-QAM), utilizing standard SER approximations; %
\(M\)-ary differential phase-shift keying (M-DPSK) with non-coherent detection, employing high-SNR SER approximations; %
and differential BPSK (DBPSK), treated as a special case of M-DPSK (\(M=2\)) using similar high-SNR SER approximations.
\newline
\noindent
For simplicity $\Xi\!\bigl(i,k;\{m\}_{q=1}^{P},\{p\}_{q=1}^{P-2}\bigr) = \Xi$,
By substituting the expression from Eq. \eqref{eq:4} into Eq. \eqref{eq:5}, CDF for the SNR is obtained,
\begin{equation}
\begin{aligned}
F_{\Upsilon}(y)
&= \sum_{i=1}^{P}\sum_{k=1}^{m_{i}}
   \Xi\times\Bigl[1-\exp\!\Bigl(
       -\frac{y}{\mu_{i}}
       \sum_{\ell=0}^{m_{i}-1}
       \frac{1}{\ell!}
       \Bigl(\frac{y}{\mu_{i}}\Bigr)^{\ell}
   \Bigr)
   \Bigr].
\end{aligned}
\end{equation}
We can also write,
\begin{equation}
\begin{aligned}
F_{\Upsilon}(y)
&= \sum_{i=1}^{P}\sum_{k=1}^{m_{i}}\Xi
 -\sum_{i=1}^{P}\sum_{k=1}^{m_{i}}
   \Xi\,
   \exp\!\biggl(
      -\frac{y}{\mu_{i}}\sum_{\ell=1}^{m_{i}-1}
      \frac{1}{\ell!}
      \left(\frac{y}{\mu_{i}}\right)^{\ell}
   \biggr).
\end{aligned}
\label{eq:F_gamma_distributed}
\end{equation}
By substituting the expression from \eqref{eq:F_gamma_distributed} into \eqref{eq:MRC_SER}, CDF for the SNR is obtained,
\begin{equation}
\begin{aligned}
\overline{P_e}
&= \frac{\mathcal{A}\sqrt{\mathcal{B}}}{2\sqrt{\pi}}
   \int_{0}^{\infty} y^{-1/2} e^{-\mathcal{B}y}
   \sum_{i=1}^{P}\sum_{k=1}^{m_i}
   \Xi dy \\
&\quad -\frac{\mathcal{A}\sqrt{\mathcal{B}}}{2\sqrt{\pi}}
   \int_{0}^{\infty} y^{-1/2} e^{-\mathcal{B}y}
   \sum_{i=1}^{P}\sum_{k=1}^{m_i}
   \Xi\\
   &\quad \times \exp\!\Bigl(
        -\tfrac{y}{\mu_i}
        \sum_{\ell=1}^{m_i-1}
        \tfrac{1}{\ell!}\bigl(\tfrac{y}{\mu_i}\bigr)^{\ell}
   \Bigr)\,dy.
\end{aligned}
\label{eq:Pe_bar}
\end{equation}
After some mathematical manipulations,
\begin{equation}
	\begin{aligned}
		\overline{P_e}
		&= \frac{\mathcal{A}\sqrt{\mathcal{B}}}{2\sqrt{\pi}}
		\sum_{i=1}^{P} \sum_{k=1}^{m_{i}}
		\Xi\int_{0}^{\infty} 
		y^{-\tfrac{1}{2}}\,
		e^{-\mathcal{B} y}dy -\frac{\mathcal{A}\sqrt{\mathcal{B}}}{2\sqrt{\pi}}
		\sum_{i=1}^{P} \sum_{k=1}^{m_{i}}\sum_{\ell=1}^{m_{i}-1}\\
        &\quad \Xi\times
		\left(\frac{1}{\ell!}\right)\int_{0}^{\infty} 
		y^{-\tfrac{1}{2}}\,
		e^{-\mathcal{B} y}
		\,e^{-\frac{y}{\mu_{i}}}
		\left( \frac{y}{\mu_{i}} \right)^{\ell}
		dy.
	\end{aligned}
		\label{eq:Pe_bar2}
\end{equation}
Here, in \eqref{eq:Pe_bar2} the integrals \(\mathcal{C}\) and \(\mathcal{D}\) are defined as follows:
\begin{subequations}
	\label{eq:CD_definitions}
	\begin{align}
		\mathcal{C} &= \int_{0}^{\infty} 
		y^{-\tfrac{1}{2}}\,
		e^{-\mathcal{B} y}\, dy, \label{eq:C_def} \\[1mm]
		\mathcal{D} &= \int_{0}^{\infty} 
		y^{-\tfrac{1}{2}}\,
		e^{-\mathcal{B} y}\,
		e^{-\frac{y}{\mu_{i}}}\,
		\left( \frac{y}{\mu_{i}} \right)^{\ell}\, dy.
		\label{eq:D_def}
	\end{align}
\end{subequations}
We can write,
\begin{equation}
	\mathcal{C} \;=\;
	\int_{0}^{\infty} y^{-\frac{1}{2}} e^{-\mathcal{B} y} \, dy 
	\;=\;
	\sqrt{\frac{\pi}{\mathcal{B}}},
	\quad \mathcal{B} > 0,
	\label{eq:C_integral}
\end{equation}
\noindent
where the result follows from \cite[Eq.~3.362.2]{Gradshteyn:1702455}.
Similarly,
	\begin{equation}
		\begin{aligned}
			\mathcal{D} \;=\;& \int_{0}^{\infty}
			y^{-\tfrac{1}{2}}\, e^{-\mathcal{B} y}\,
			e^{-\tfrac{y}{\mu_i}}\,
			\Bigl(\tfrac{y}{\mu_i}\Bigr)^{\ell} \, dy 
			\\[6pt]
			=\;& \left(\frac{1}{\mu_i}\right)^{\ell}
			\frac{\Gamma\!\Bigl(\ell + \tfrac{1}{2}\Bigr)}
			{\Bigl(\mathcal{B} + \tfrac{1}{\mu_i}\Bigr)^{\ell + \tfrac{1}{2}}}
			\quad \text{for } \operatorname{Re}\Bigl(\mathcal{B} + \frac{1}{\mu_i}\Bigr) > 0,
			\\[8pt]
			=\;& \left(\frac{1}{\mu_i}\right)^{\ell}
			\frac{(2\ell - 1)!!\,\sqrt{\pi}}
			{2^{\ell}\,\Bigl(\mathcal{B} + \tfrac{1}{\mu_i}\Bigr)^{\ell + \tfrac{1}{2}}}
			\quad \text{for } \operatorname{Re}\Bigl(\mathcal{B} + \frac{1}{\mu_i}\Bigr) > 0.
		\end{aligned}
		\label{eq:D_integral_doublefactorial}
	\end{equation}
\noindent
This result follows from applying standard Gamma-function integrals, specifically \cite[Eq.~2.3.3.3]{Prudnikov_Brychkov_Marichev_Queen_1986}. 
\newline
\noindent
Finally, for single-user case BER can be derived from the SER expression,
\begin{equation}
\bar{P_b} = \frac{\mathcal{A}}{2 \text{log}_2\mathcal{M}}\left[ \sum_{i=1}^{P}\sum_{k=1}^{m_i}\Xi \times\mathcal{C}-\sum_{i=1}^{P}\sum_{k=1}^{m_i}\sum_{l=1}^{m_i-1}\left( \frac{1}{l!} \right)\Xi \times\mathcal{D}\right].
\end{equation}

\subsection{Multi-User Case}
In a SIMO system with $K_u$ users and $P$ diversity paths, the signal-to-interference-plus-noise ratio (SINR) for the $k$-th user is defined based on the energy of the desired signal, noise, and the interference caused by the remaining users. Assuming that the interference originates from users $k' = k+1$ to $K_u$, the SINR expression becomes \cite{9814545},
\begin{equation}
	\Upsilon_k \triangleq \frac{E_s}{N_0 + \sigma_k^2}
	= \frac{E_s}{N_0 + E_s \sum_{k'=k+1}^{K_u} \sum_{p=1}^P \bigl\lvert h_p^{(k')} \bigr\rvert^2}.
\end{equation}
We first define the SINR-related variable \cite{ilhan2015performance},
\begin{equation}
	\Upsilon_k \;=\; \frac{Y_k}{Z_k},
\end{equation}
where, for convenience, we also denote
\begin{equation}
	{Y_k} \;=\;\frac{E_s}{N_0}.
\end{equation}
Therefore, we can write,
\begin{equation}
	\label{eq:Zk}
	Z_k = 1 + \frac{E_s}{N_0}
	\sum_{k'=k+1}^{K_u}
	\sum_{p=1}^{P}
	\bigl\lvert h_p^{(k')}\bigr\rvert^2.
\end{equation}
Let $|h_{p}^{(k')}|^2$ denote the squared magnitude of the channel coefficient for the $p$-th path of the $k'$-th interfering user. Assuming Nakagami–\(m\) fading, each $|h_{p}^{(k')}|^2$ follows a Gamma distribution as,
\begin{equation}
    |h_{p}^{(k')}|^2 \sim \Gamma\left(m_{p}^{(k')},\; \frac{\Omega_{p}^{(k')}}{m_{p}^{(k')}}\right)
\end{equation}
where $m_{p}^{(k')}$ is shape parameter (Nakagami–\(m\) factor) for the $p$-th path of user $k'$ and $\Omega_{p}^{(k')}$ is second moment scale parameter ($\mathbb{E}[|h_{p}^{(k')}|^2] = \Omega_{p}^{(k')}$).
The aggregate interference power from $K_u-k$ interfering users, each with $P$ paths, is given by:

\begin{equation}
    S = \frac{E_s}{N_0} \sum_{k'=k+1}^{K_u} \sum_{p=1}^P |h_p^{(k')}|^2
\end{equation}
The first two moments of the interference power $S$ are derived as follows. The mean interference power is,
\begin{equation}
\mu_S = \frac{E_s}{N_0} \sum_{k'=k+1}^{K_u} \sum_{p=1}^{P} \Omega_p^{(k')}
\label{eq:mean_power}
\end{equation}
The variance of the interference power is,
\begin{equation}
\sigma_S^2 = \left(\frac{E_s}{N_0}\right)^2 \sum_{k'=k+1}^{K_u} \sum_{p=1}^{P} \frac{(\Omega_p^{(k')})^2}{m_p^{(k')}}
\label{eq:variance_power}
\end{equation}
Using the moments matching \cite{9290053}, we approximate $S$ as a Gamma-distributed random variable $S \sim \Gamma(m_z, \Omega_z)$. The shape and scale parameters are respectively,
\begin{equation}
m_z = \frac{\mu_S^2}{\sigma_S^2} = \frac{\left(\sum\limits_{k'=k+1}^{K_u} \sum\limits_{p=1}^{P} \Omega_p^{(k')}\right)^2}{\sum\limits_{k'=k+1}^{K_u} \sum\limits_{p=1}^{P} \frac{(\Omega_p^{(k')})^2}{m_p^{(k')}}}
\label{eq:shape_param}
\end{equation}
\begin{equation}
\Omega_z = \frac{\sigma_S^2}{\mu_S} = \frac{E_s}{N_0} \cdot \frac{\sum\limits_{k'=k+1}^{K_u} \sum\limits_{p=1}^{P} \frac{(\Omega_p^{(k')})^2}{m_p^{(k')}}}{\sum\limits_{k'=k+1}^{K_u} \sum\limits_{p=1}^{P} \Omega_p^{(k')}}
\label{eq:scale_param}
\end{equation}
This Gamma approximation provides a complete statistical characterization of the interference in multi-user OTFS systems with arbitrary Nakagami-$m$ fading parameters. The parameters \( Z_k = 1 + S \), where \( S \sim \Gamma(m_z, \Omega_z) \), and thus \( Z_k - 1 \sim \Gamma(m_z, \Omega_z) \), the following derivations hold, PDF of \( Z_k \) for \( z > 1 \) is given by,
\begin{equation}
    f_{Z_k}(z) = \frac{(z-1)^{m_z-1}}{\Omega^{m_z} \Gamma(m_z)} e^{-\frac{z-1}{\Omega_z}}, 
    \label{eq:pdf_zk}
\end{equation}
For \( \Upsilon_k = \frac{E_s/N_0}{Z_k} \), using the transformation \( Z_k = \frac{E_s/N_0}{\Upsilon_k} \) and \( \left|\frac{dz_k}{d\Upsilon_k}\right| = \frac{E_s/N_0}{\Upsilon_k^2} \), the PDF of \( \Upsilon_k \) becomes,
\begin{equation}
f_{\Upsilon_k}(y) = \frac{\left( \frac{E_s}{N_0 y} - 1 \right)^{m_z - 1}}{\Omega_z^{m_z} \Gamma(m_z)} \cdot \frac{E_s / N_0}{y^2} \exp\left( -\frac{\frac{E_s}{N_0 y} - 1}{\Omega_z} \right),
\end{equation}
where $0 \leq y \leq \frac{E_s}{N_0}$.
The CDF is obtained by integrating the PDF,
\begin{equation}
    F_{\Upsilon_k}(y) = \int_0^{y} f_{\Upsilon_k}(t) \, dt  = \frac{1}{\Gamma(m_z)} \, \gamma\left( m_z, \frac{\frac{E_s}{N_0 y} - 1}{\Omega_z} \right),
\label{eq:cdf_gammak}
\end{equation}
where \( \gamma(\cdot, \cdot) \) denotes the lower incomplete Gamma function. By substituting the expression from Eq. \eqref{eq:cdf_gammak} into Eq. \eqref{eq:MRC_SER},
invoking the identities \cite[Eq.~8.4.16.1]{Prudnikov_Brychkov_Marichev_Queen_1986} and \cite[Eq.~8.4.16.2]{Prudnikov_Brychkov_Marichev_Queen_1986}, the lower and upper incomplete Gamma functions can be expressed in closed form through the Meijer-\(G\) function as follows,
\begin{equation}
  \gamma\!\left(m_z,\frac{E_s/N_0 y-1}{\Omega_z}\right)
  \;=\;
  \Gamma(m_z)
  -\Gamma\!\left(m_z,\frac{E_s/N_0 y-1}{\Omega_z}\right),
\end{equation}
\begin{equation}
  \Gamma\!\left(m_z,\frac{E_s/N_0 y-1}{\Omega_z}\right)
  \;=\;
  G^{2,0}_{1,2}\!\Bigl(
      \tfrac{E_s/N_0 y-1}{\Omega_z}
      \,\Big|\,
      \begin{array}{c} 1 \\ m_z,0 \end{array}
  \Bigr).
\end{equation}
The exponential term is converted to Meijer-$G$ form \cite{Mathai_Saxena_1978}:
\begin{equation}
	\exp\left(-\left(\frac{E_s}{N_0 y} - 1\right)\right) = G_{0,1}^{1,0}\left( \left(\frac{E_s}{N_0 y} - 1\right) \left| \begin{array}{c} - \\ 0 \end{array} \right. \right)
\end{equation}
The average SER can therefore be expressed in terms of Meijer-$G$ functions as the following
single-integral representation,
\begin{equation}
\begin{aligned}
  \overline{P}_{e}
  &= \frac{\mathcal{A}}{2}\sqrt{\frac{\mathcal{B}}{\pi}}
     \int_{0}^{\infty}
       y^{-1/2}\,
       G^{2,0}_{1,2}\!\Bigl(
         \tfrac{E_s/N_0 y- 1}{\Omega_z}
         \,\Big|\,
         \begin{array}{c} 1 \\ m_z,0 \end{array}
       \Bigr) \\[4pt]
  &\quad \times
       G^{1,0}_{0,1} \left( \frac{E_s}{N_0 y} - 1 \,\middle|\, \begin{array}{c}
- \\
0
\end{array} \right)\times e^{-\mathcal{B}y}dy .
\end{aligned}
\label{eq:BER_MRC_integral}
\end{equation}
Invoking the product–integration formula for Meijer–\(G\) functions
\cite[Eq.~2.24.1.1]{Prudnikov_Brychkov_Marichev_Queen_1986} and 
\eqref{eq:C_integral} the following closed-form relation of SER is obtained,
\begin{equation}
  \overline{P}_{e}
  \;=\;
  \frac{\mathcal{A}}{2}\,
  G^{3,1}_{2,3}\!\Bigl(
     \tfrac{E_s}{\Omega_z N_0}
     \,\Big|
     \begin{array}{c}
       1-m_z,\;1 \\[2pt]
       0,\;1-m_z,\;\tfrac12
     \end{array}
  \Bigr)
\end{equation}
In the multi-user case, BER expression is derived as,
\begin{equation}
  \overline{P}_{b}
  \;=\;
  \frac{\mathcal{A}}{2\log_2\mathcal{M}}\,
  G^{3,1}_{2,3}\!\Bigl(
     \tfrac{E_s}{\Omega_z N_0}
     \,\Big|
     \begin{array}{c}
       1-m_z,\;1 \\[2pt]
       0,\;1-m_z,\;\tfrac12
     \end{array}
  \Bigr).
\end{equation}
\section{Numerical Results}
\label{sec:numerical_results}
This section presents the analysis of the BER performance for OTFS modulation systems under Nakagami–\(m\) fading channel conditions. We examine both single-path ($P=1$) and multi-path ($P=2$) scenarios, using BPSK and QPSK modulation across different fading parameters. The numerical results are obtained through Monte Carlo simulations that implement both our theoretical derivations and ML detection. In all simulations, the Extended Vehicular A (EVA) channel model is adopted to represent realistic time-dispersive and Doppler-rich wireless channel conditions \cite{3gpp_ts_36_101}. The simulation parameters used in our analysis are summarized in Table \ref{table_2}.
\begin{table}[ht]
\centering
\caption{Simulation Parameters}
\label{tab:sim_params}
\begin{threeparttable}
\begin{tabular*}{\columnwidth}{@{\extracolsep{\fill}} l l @{}}
\toprule
Parameter & Value\\
\midrule
Detection method & ML  \\
Modulation schemes & BPSK, QPSK  \\
Doppler bins ($N$) & 2 \\
Delay bins ($M$) & 2 \\
Number of propagation paths ($P$) & 1, 2 \\
Nakagami–\(m\) parameter ($m$) & 1, 2 \\
Number of receive users ($K_u$) & 2 \\
Carrier frequency ($f_c$) & 4 GHz \\
Subcarrier spacing ($\Delta f$) & 15 kHz \\
Maximum user speed & 120 km/h \\
Channel model & EVA \\
\bottomrule
\end{tabular*}
\end{threeparttable}
\label{table_2}
\end{table}
\subsection{SISO-OTFS Performance Analysis}
\begin{figure}[h]
    \centering
    \includegraphics[width=1\linewidth]{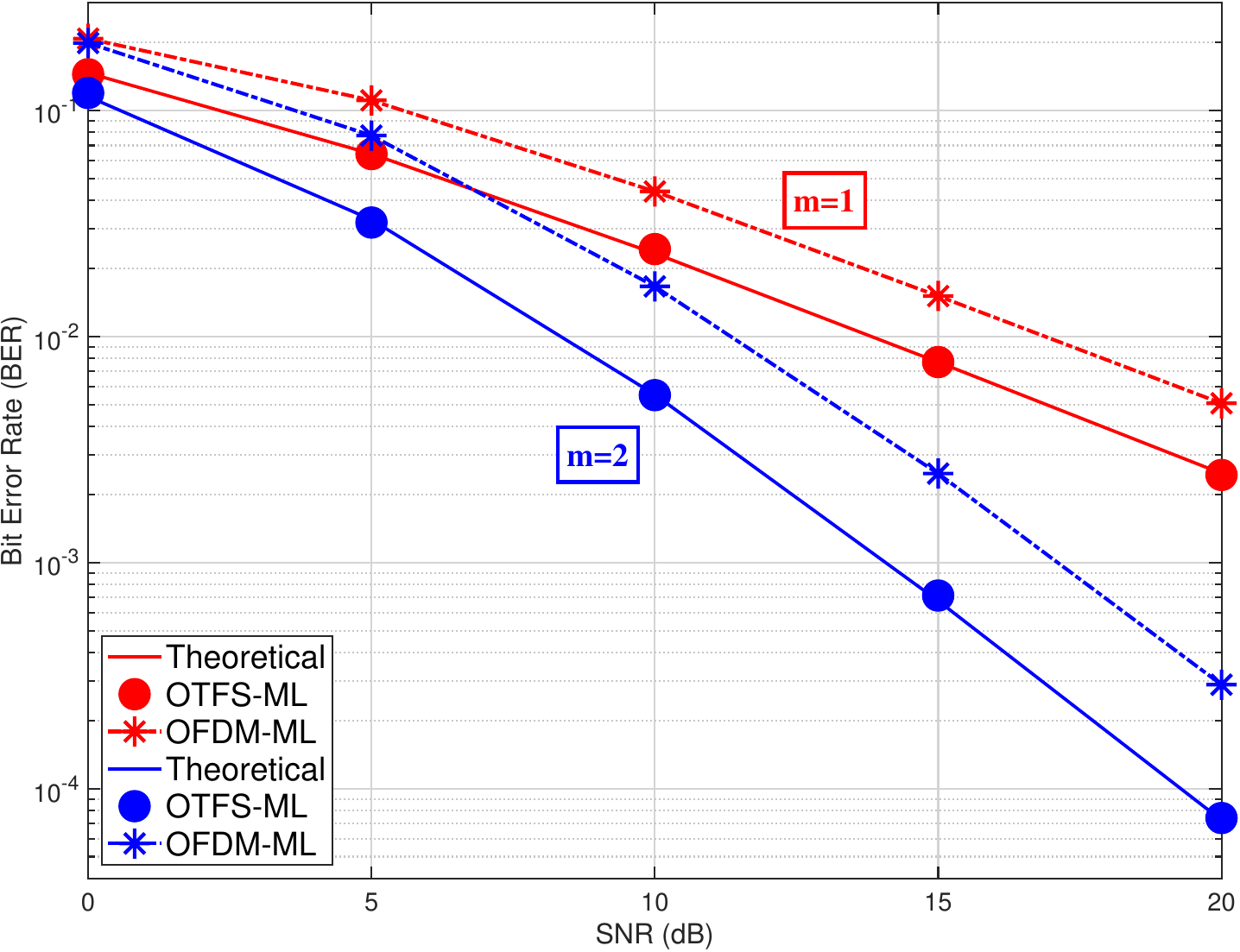}
    \caption{BER performance of SISO-OTFS and OFDM systems employing BPSK modulation over Nakagami-$m$ fading channels with $P=1$ and $m=1,2$.}
    \label{fig:1}
\end{figure}

\begin{figure}
    \centering
    \includegraphics[width=1\linewidth]{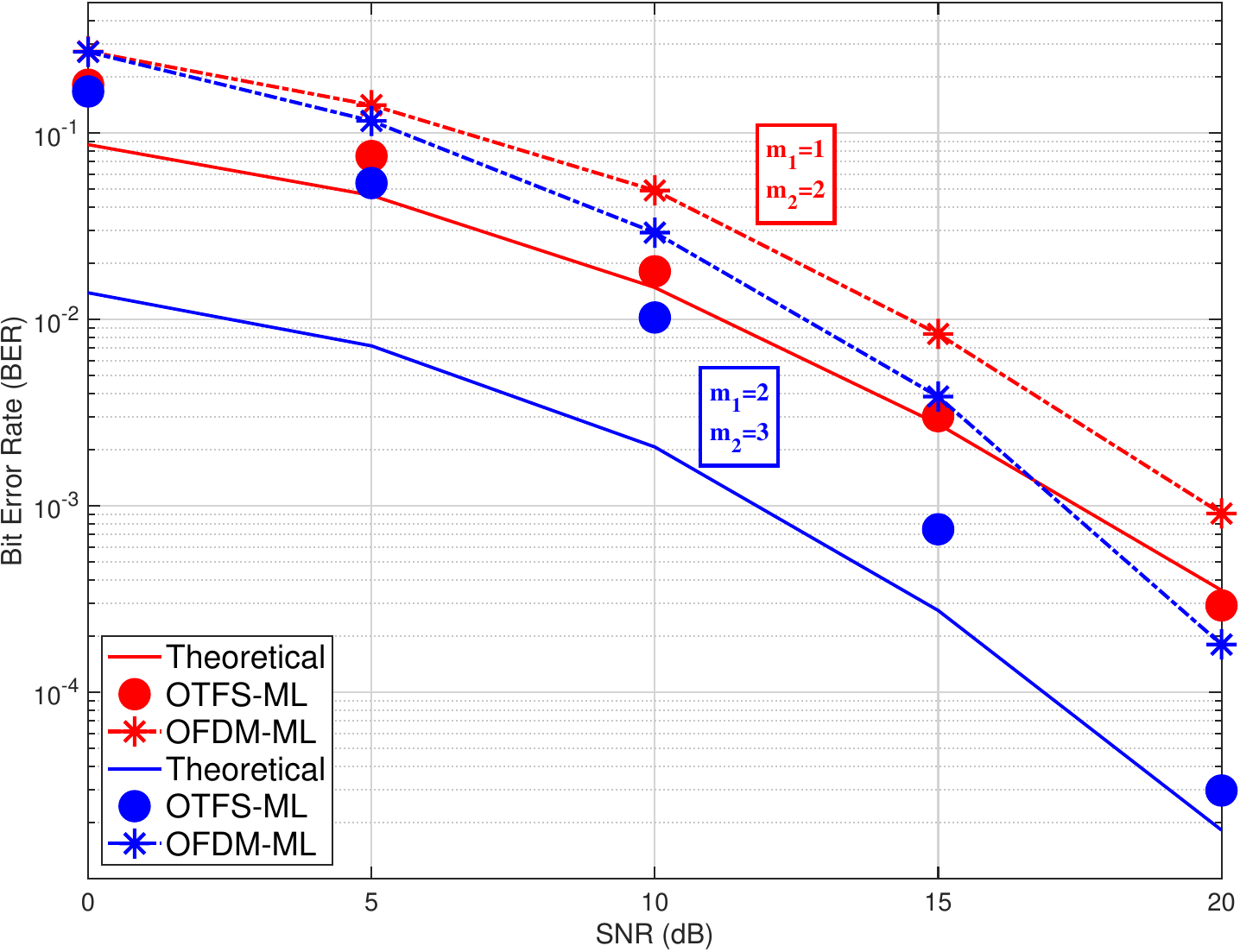}
    \caption{BER performance of SISO-OTFS and OFDM systems employing QPSK modulation over Nakagami-$m$ fading channels with double path ($P=2$) for different fading parameters $(m_1, m_2)$.}
    \label{fig:2}
\end{figure}

Fig. \ref{fig:1} and Fig. \ref{fig:2} illustrate BER performance of SISO-OTFS systems under varying channel conditions with different number of propagation paths $P$ and Nakagami fading parameter $m$ values. The theoretical curves are computed using the derived $\overline{P_b}$ function for the error probability by using Erlang CDF.

Fig.~\ref{fig:1} illustrates BER performance of SISO-OTFS and conventional OFDM systems employing BPSK modulation over Nakagami-$m$ fading channels for $m=1$ and $m=2$. Both empirical (OTFS-ML) and theoretical OTFS results, as well as OFDM-ML simulation outcomes, are depicted as a function of SNR. The results reveal a close agreement between OTFS-ML simulations and the theoretical predictions for both fading severities, confirming the validity of the analytical framework. For $m=1$ (Rayleigh fading) at 20 dB SNR, OTFS-ML achieves a BER of $2.44 \times 10^{-3}$, while OFDM-ML yields $5.06 \times 10^{-3}$, demonstrating a clear performance improvement for OTFS. With $m=2$, representing milder fading, the BER values for both OTFS and OFDM decrease, and OTFS-ML consistently outperforms OFDM-ML, achieving a BER of $7.4 \times 10^{-5}$ at 20 dB SNR. These results demonstrate that OTFS not only provides better performance than conventional OFDM—particularly in severe fading ($m=1$)—but also that the theoretical analysis remains accurate for different Nakagami-$m$ values. The diversity gain enhancement due to higher $m$ values is evident from the steeper BER curves, indicating improved robustness of both OTFS and OFDM as the channel fades less severely.

Fig. ~\ref{fig:2} compares the BER performance of SISO-OTFS and conventional OFDM systems employing QPSK modulation in double-path ($P=2$) Nakagami-$m$ fading channels for two different fading parameter sets: $(m_1, m_2) = (1, 2)$ and $(m_1, m_2) = (2, 3)$. In the analytical evaluation of OTFS, due to the definition of the $\Xi$ expression, the shape parameters $m_1$ and $m_2$ for each path are set as distinct integer values with $m_2 > m_1$, and the average path powers ($\mu_1,\ \mu_2$) are also taken to be different. Both analytical results for OTFS (Theoretical) and simulation results based on maximum likelihood detection (OTFS-ML and OFDM-ML) are presented as a function of SNR. For $(m_1, m_2) = (1,2)$ at 20 dB SNR, OTFS-ML achieves a BER of $2.92 \times 10^{-4}$, significantly improving upon the $9.08 \times 10^{-4}$ BER offered by OFDM-ML. When the fading conditions are further alleviated to $(m_1, m_2) = (2,3)$, the performance benefit of OTFS-ML is further enhanced, with BER dropping to $2.97 \times 10^{-5}$ versus $1.80 \times 10^{-4}$ for OFDM-ML at the same SNR. In all cases, OTFS-ML results show excellent agreement with the theoretical curves, confirming the validity of the analytical framework. The improvement in BER and the increased slope of the OTFS curves as the Nakagami fading parameters rise clearly demonstrate the diversity gain advantage enabled by higher $m$ values and the double-path scenario. These results confirm that OTFS consistently outperforms OFDM in terms of error resilience, especially as the combined effects of multipath and reduced fading severity are considered.
\subsection{SIMO-OTFS Performance Analysis}

\begin{figure}
    \centering
    \includegraphics[width=1\linewidth]{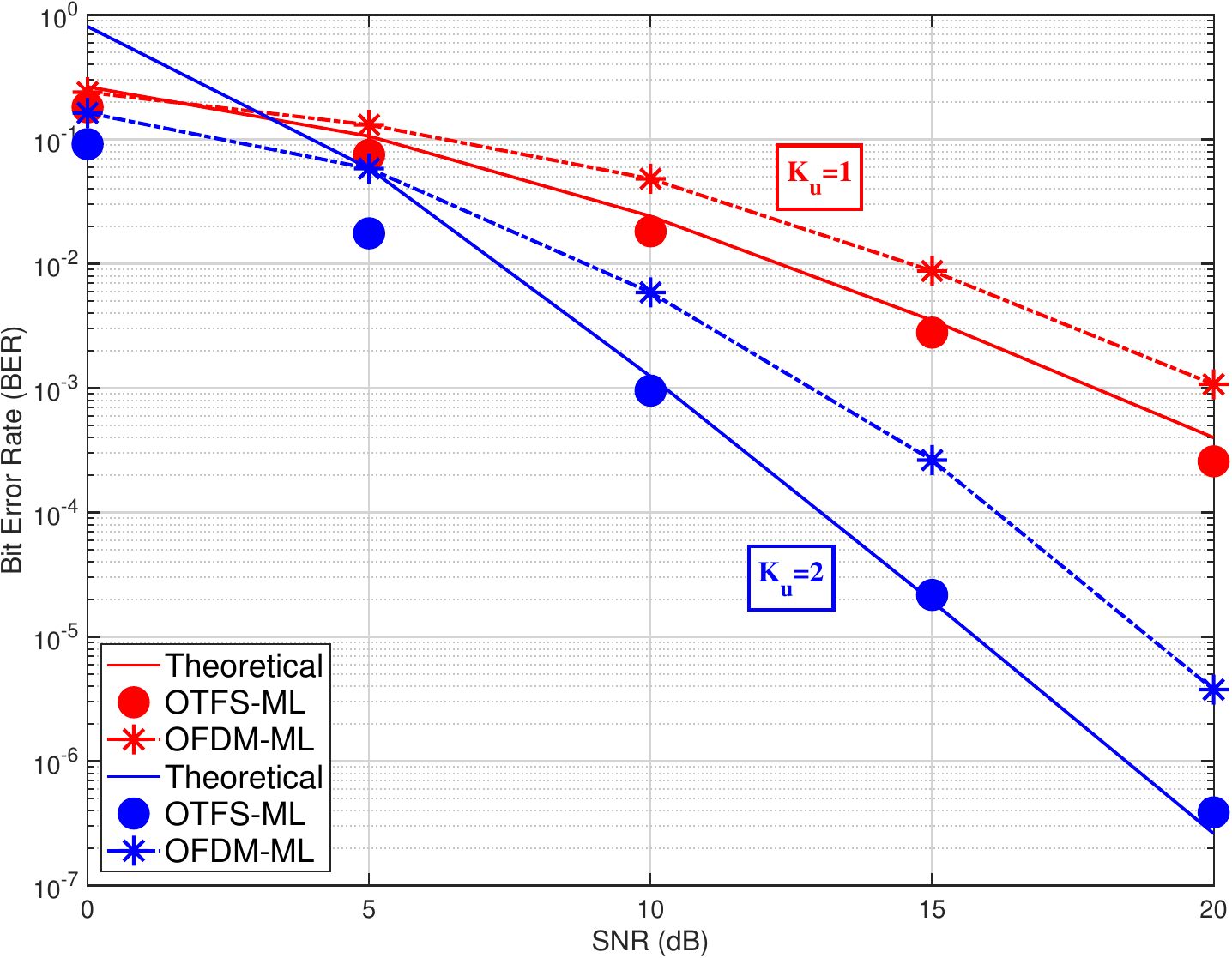}
    \caption{BER performance of SIMO-OTFS and OFDM systems with QPSK modulation over Nakagami-$m$ fading channels for $P=1$, $m=2$, and different numbers of users ($K_u=1,2$).}
    \label{fig:3}
\end{figure}

\begin{figure}
    \centering
\includegraphics[width=1\linewidth]{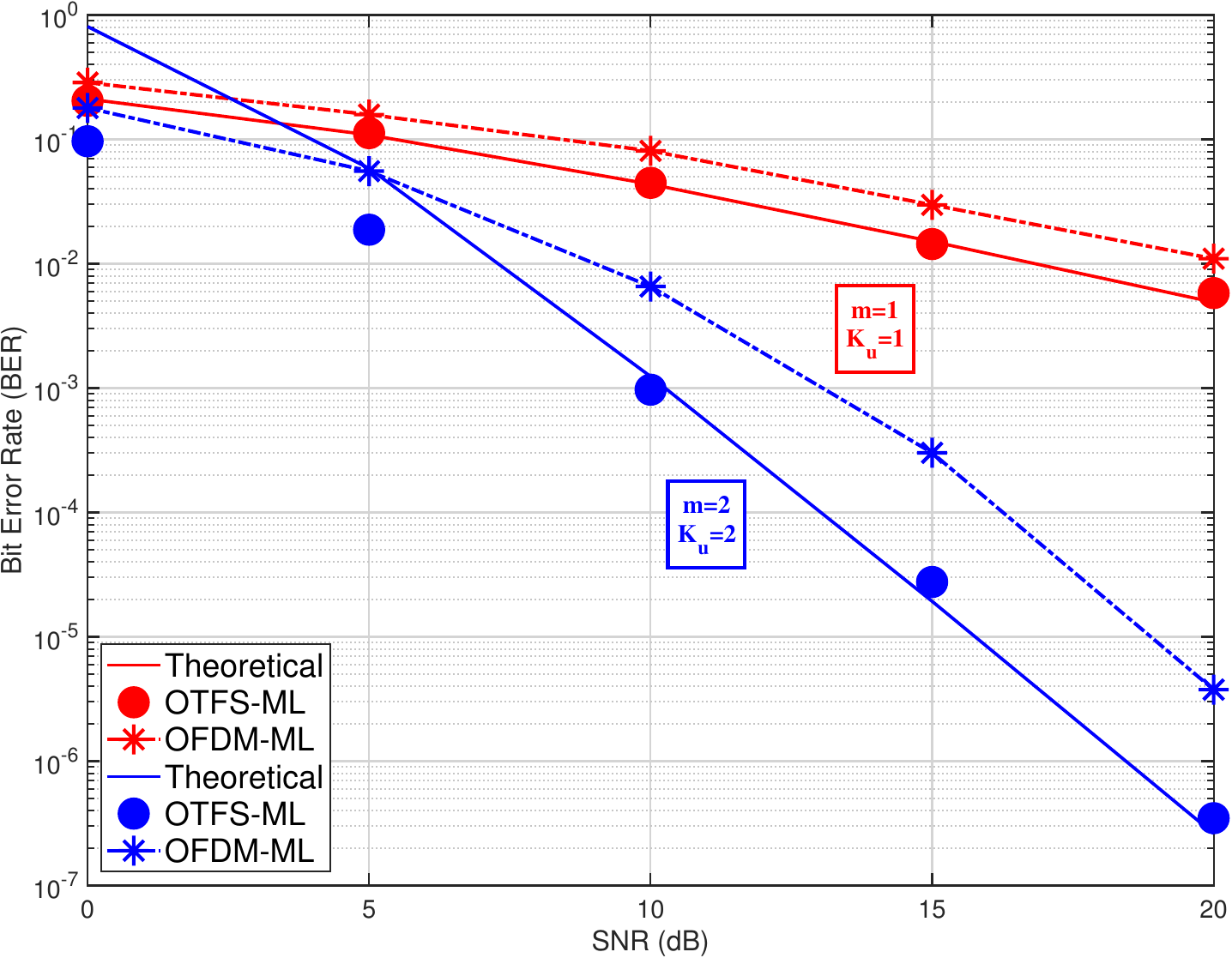}
    \caption{BER performance of SIMO-OTFS and OFDM systems with QPSK modulation over two-path ($P=2$) Nakagami-$m$ fading channels.}
    \label{fig:4}
\end{figure}

The BER performance of the SIMO OTFS system is illustrated in Fig. ~\ref{fig:3} for single-path ($P=1$) Nakagami-$m$ channels with $m=2$ and varying user numbers ($K_u=1,2$), and in Fig. ~\ref{fig:4} for two-path ($P=2$) channels that contrast scenarios of severe fading with $m=1, K_u=1$ and milder fading with $m=2, K_u=2$. These analyses focus on the influence of increasing the number of users as well as improving the fading distribution, with all cases employing QPSK modulation. Theoretical BER curves for OTFS are generated via moment matching and expressed in terms of Meijer-$G$ functions to accurately capture the effects of multi-user interference. The theoretical results are benchmarked against simulation outcomes using maximum likelihood detection for both OTFS and OFDM systems. This comprehensive evaluation captures how spatial diversity and channel conditions impact system robustness.

Fig.~\ref{fig:3} presents the BER performance of SIMO-OTFS and conventional OFDM systems for single-path ($P=1$), QPSK modulation, and Nakagami-$m$ fading with $m=2$. The figure compares systems with one user ($K_u=1$) and two users ($K_u=2$), showing both theoretical predictions and simulation outcomes (OTFS-ML and OFDM-ML) as a function of SNR. With $K_u=1$ at 20 dB SNR, OTFS-ML achieves a BER of $2.58 \times 10^{-4}$, outperforming OFDM-ML, which yields $1.07 \times 10^{-3}$. When the number of users increases to $K_u=2$, a further BER reduction is observed for OTFS-ML, reaching $3.87 \times 10^{-7}$ at 20 dB, while OFDM-ML’s BER is $3.77 \times 10^{-6}$. In both cases, the analytical and simulation results for OTFS are in excellent agreement. Notably, increasing the number of users provides a substantial diversity gain, reflected in the steeper BER reduction of the curves, and further highlights the robustness of OTFS compared to OFDM under favorable channel conditions and multi-user operation.

Fig.~\ref{fig:4} presents the BER results for SIMO-OTFS and OFDM systems with QPSK, $P=2$ and two representative configurations: severe fading with $(m=1;K_u=1)$ and milder fading with $(m=2; K_u=2)$. Both ML simulation and theoretical curves for OTFS are shown together with OFDM-ML results. For the $(m=1;K_u=1)$, at 20 dB SNR, OTFS-ML provides a BER of $5.81\times10^{-3}$, compared to $1.10\times10^{-2}$ for OFDM-ML. When $m=2$ and $K_u=2$, the error rates drop sharply and OTFS-ML achieves $3.48\times10^{-7}$, while OFDM-ML records $3.76\times10^{-6}$. The simulation and analytical results for OTFS match well in both regimes. The evident drop in BER when either the fading conditions improve or more users/antennas are employed, highlights OTFS’s ability to effectively harness the available diversity. These trends underline the robust performance of OTFS, which efficiently exploits enhanced diversity stemming from multi-paths and spatial/user resources.
\subsection{Diversity Gain Analysis}
Diversity gain quantifies the robustness of a wireless communication system to fading, measuring how rapidly the BER decreases as the SNR increases, especially at high SNR. Formally, the diversity gain $G_d$ is defined as the negative slope of the BER curve on a log-log scale with respect to SNR \cite{DIV_GAIN},
\begin{equation}
G_d = -\frac{\partial \log_{10} \overline{P_b}}{\partial \log_{10} \Upsilon}\bigg|_{\Upsilon \to \infty}
\label{eq:diversity_gain}
\end{equation}
For the SISO case, consider a system employing QPSK modulation and Nakagami-$m$ fading with various numbers of independent multipath components, denoted by $P$. The (approximate) empirical diversity gain between two SNR points $\Upsilon_1$ and $\Upsilon_2$ can be calculated as
\begin{equation}
G_d \approx - \frac{ \log \overline{P_{b,2}} - \log \overline{P_{b,1}} }{ \log \Upsilon_2 - \log \Upsilon_1 }
\label{eq:slope_discrete}
\end{equation}
where $\overline{P_{b,1}}$ and $\overline{P_{b,2}}$ denote the simulated BER at $\Upsilon_1$ and $\Upsilon_2$, respectively. The diversity gain for both SISO and SIMO systems are computed and summarized as follows, allowing direct comparison of the diversity benefits in each configuration.
\newline
\noindent
\underline{\textbf{SISO case:}}
\begin{itemize}
\item \textit{QPSK, $m=1$, $P=1$ (OTFS, simulation):}
\begin{align*}
\overline{P_{b,1}} &= 0.0441 \quad \text{at SNR} = 10 \text{ dB} \\
\overline{P_{b,2}} &= 0.00477 \quad \text{at SNR} = 20 \text{ dB} \\
G_d^{\text{OTFS}} &\approx -\frac{\log (0.00477) - \log (0.0441)}{\log (100) - \log (10)} \approx 1.0
\end{align*}
For conventional OFDM under the same conditions:
\begin{align*}
\overline{P_{b,1}}^{\text{OFDM}} &= 0.07919 \quad \text{at SNR} = 10 \text{ dB} \\
\overline{P_{b,2}}^{\text{OFDM}} &= 0.00903 \quad \text{at SNR} = 20 \text{ dB} \\
G_d^{\text{OFDM}} &\approx -\frac{\log (0.00903) - \log (0.07919)}{\log (100) - \log (10)} \approx 0.94
\end{align*}
Thus, OTFS achieves a slightly higher diversity gain compared to OFDM, indicating improved robustness against fading.

\item \textit{QPSK, $m_1=1$, $m_2=2$, $P=2$:}
\begin{align*}
\overline{P_{b,1}} &= 0.0181, \quad \overline{P_{b,2}} = 0.0002918 \\
G_d^{\text{OTFS}} &\approx -\frac{\log_{10} (0.0002918) - \log_{10} (0.0181)}{1.0} \approx 1.78
\end{align*}
\begin{align*}
\overline{P_{b,1}}^{\text{OFDM}} &= 0.04907 \quad \text{at SNR} = 10 \text{ dB} \\
\overline{P_{b,2}}^{\text{OFDM}} &= 0.000908 \quad \text{at SNR} = 20 \text{ dB}\\
G_d^{\text{OFDM}} &\approx -\frac{\log (0.000908) - \log (0.04907)}{1.0} \approx 1.73
\end{align*}
\end{itemize}

Here, increasing the number of independent paths $P$ leads to a significant increase in the system's diversity order for both OTFS and OFDM, though OTFS preserves a marginal advantage. These results confirm that the empirical diversity gain increases with the number of independent multipath components and that OTFS provides slightly higher diversity compared to OFDM under identical channel configurations. This observation is consistent with OTFS's intrinsic exploitation of delay-Doppler diversity.

In SIMO or multi-user scenarios, the presence of additional antennas or users further increases potential diversity gain. However, in multi-user settings, co-channel interference may limit the achievable diversity.
\newline
\noindent
\underline{\textbf{SIMO case:}}
\begin{itemize}
    \item \textit{QPSK, $m=1$, $P=1$, $K_u=2$}:
    \begin{align*}
    & \overline{P_{b,1}} = 0.00634, \; \overline{P_{b,2}} = 0.0000693\\
    & G_d^{\text{OTFS}} \approx -\frac{\log (0.0000693) - \log (0.00634)}{2.0 - 1.0} \approx 1.97
    \end{align*}
    \item \textit{QPSK, $m=1$, $P=1$, $K_u=2$}:
    \begin{align*}
    & \overline{P_{b,1}} = 0.016385, \; \overline{P_{b,2}} = 0.0002885\\
    & G_d^{\text{OFDM}} \approx -\frac{\log (0.0002885) - \log (0.016385)}{2.0 - 1.0} \approx 1.75
    \end{align*}
\end{itemize}
Table~\ref{tab:diversity_gain_qpsk_updated} summarizes the estimated diversity gains ($G_d$) achieved for both OTFS and OFDM modulation schemes under several configurations, based on BER measurements at 10 dB and 20 dB SNR.
\begin{table}[h]
\centering
\caption{Estimated diversity gain values ($G_d$) for various configurations using BER at 10 and 20 dB SNR.}
\label{tab:diversity_gain_qpsk_updated}
\footnotesize
\renewcommand{\arraystretch}{1.5} 
\begin{tabular}{|c|c|c|} 
\hline
\textbf{Configuration} & \textbf{$G_d^{\mathrm{OTFS}}$} & \textbf{$G_d^{\mathrm{OFDM}}$} \\
\hline
SISO, $P{=}1$, $m{=}1$                   & 0.97  & 0.94  \\
SISO, $P{=}1$, $m{=}2$                   & 1.80  & 1.62  \\
SISO, $P{=}2$, $m_1{=}1$, $m_2{=}2$       & 1.79  & 1.73  \\
SISO, $P{=}2$, $m_1{=}2$, $m_2{=}3$       & 2.54  & 2.21  \\
SIMO, $P{=}1$, $m{=}1$, $K_u{=}2$         & 1.96  & 1.75  \\
SIMO, $P{=}1$, $m{=}2$, $K_u{=}2$         & 3.39  & 3.19  \\
SIMO, $P{=}2$, $m{=}2$, $K_u{=}2$         & 3.45  & 3.24  \\
\hline
\end{tabular}
\end{table}

For interference-free SIMO, the diversity order approaches the sum of the diversity provided by independent paths and antennas, as analytically stated in Eq.~(\ref{eq:diversity_gain}) and shown in theoretical works such as~\cite{DIV_GAIN}. In the presence of interference (higher $K_u$), diversity gain may degrade depending on the interference statistics. Overall, the diversity order consistently increases with the number of independent paths $P$ and antennas/users $K_u$, but may be limited by correlated channels or strong multi-user interference.

\noindent
\noindent
The variable \(m_i\) represents the Nakagami-\(m\) parameter associated with the \(i\)th diversity path. For a SISO system employing \(P\) independent diversity paths, the overall diversity gain can be approximated based on the analysis described in \cite{DIV_ILHAN} as,
\begin{equation}
G_d^{\mathrm{SISO}} = P \cdot \min_i(m_i).
\label{eq:siso_diversity_gain_compact}
\end{equation}

\noindent
The parameter \(m_{j,p}\) denotes the Nakagami-\(m\) fading parameter associated with the \(p\)th path in the \(j\)th diversity branch, where \(j = 1, \dots, K_u\) and \(p = 1, \dots, P\). In SIMO systems, where each receive branch may observe multiple independent diversity paths, the overall diversity gain can be approximated by,

\begin{equation}
\begin{aligned}
G_d^{\mathrm{SIMO}} = &\, K_u \cdot \Bigg[ \min_p(m_{j,p}) \\
&+ \frac{\log_2(1 + P)}{P} 
\sum_{p=1}^{P} \left(m_{j,p} - \min_p(m_{j,p})\right) \Bigg].
\end{aligned}
\label{eq:simo_diversity_gain_aligned}
\end{equation}

\section{Conclusion}
\label{sec:conclusion}
This paper has systematically investigated the BER performance of OTFS modulation over Nakagami-$m$ fading channels, addressing both single-user and multi-user scenarios. Key contributions include the derivation of closed-form BER expressions for SISO configurations using the Erlang probability density function, and the development of an analytical framework for SIMO systems employing moment matching and Meijer-$G$ functions to evaluate BER under co-channel interference. The impact of single-path versus multipath conditions and varying Nakagami-$m$ fading parameters was also examined. In addition, a comprehensive empirical diversity gain analysis was performed, demonstrating that OTFS attains a consistently higher diversity order than conventional OFDM as the number of independent channel paths or spatial/user resources increases, further enhancing its robustness to fading. Our analytical derivations were validated through Monte Carlo simulations with MLD, confirming a strong correlation with numerical results. Crucially, comparative evaluations demonstrated that OTFS consistently achieves superior error performance over conventional OFDM in the investigated high-mobility Nakagami-$m$ fading scenarios for both BPSK and QPSK modulations. This study underscores the enhanced resilience of OTFS to challenging channel dynamics.
The inherent robustness of OTFS modulation, particularly its efficacy in high-mobility and dispersive channel conditions, positions it as a compelling candidate waveform for 6G and future wireless communication systems.
\vspace{2pt}
\bibliographystyle{IEEEtran}
\bibliography{Referanslar}
 \newpage
\section{Biography Section}
 \vspace{-33pt} 
\begin{IEEEbiography}[{\includegraphics[width=1in,height=1.25in,clip,keepaspectratio]{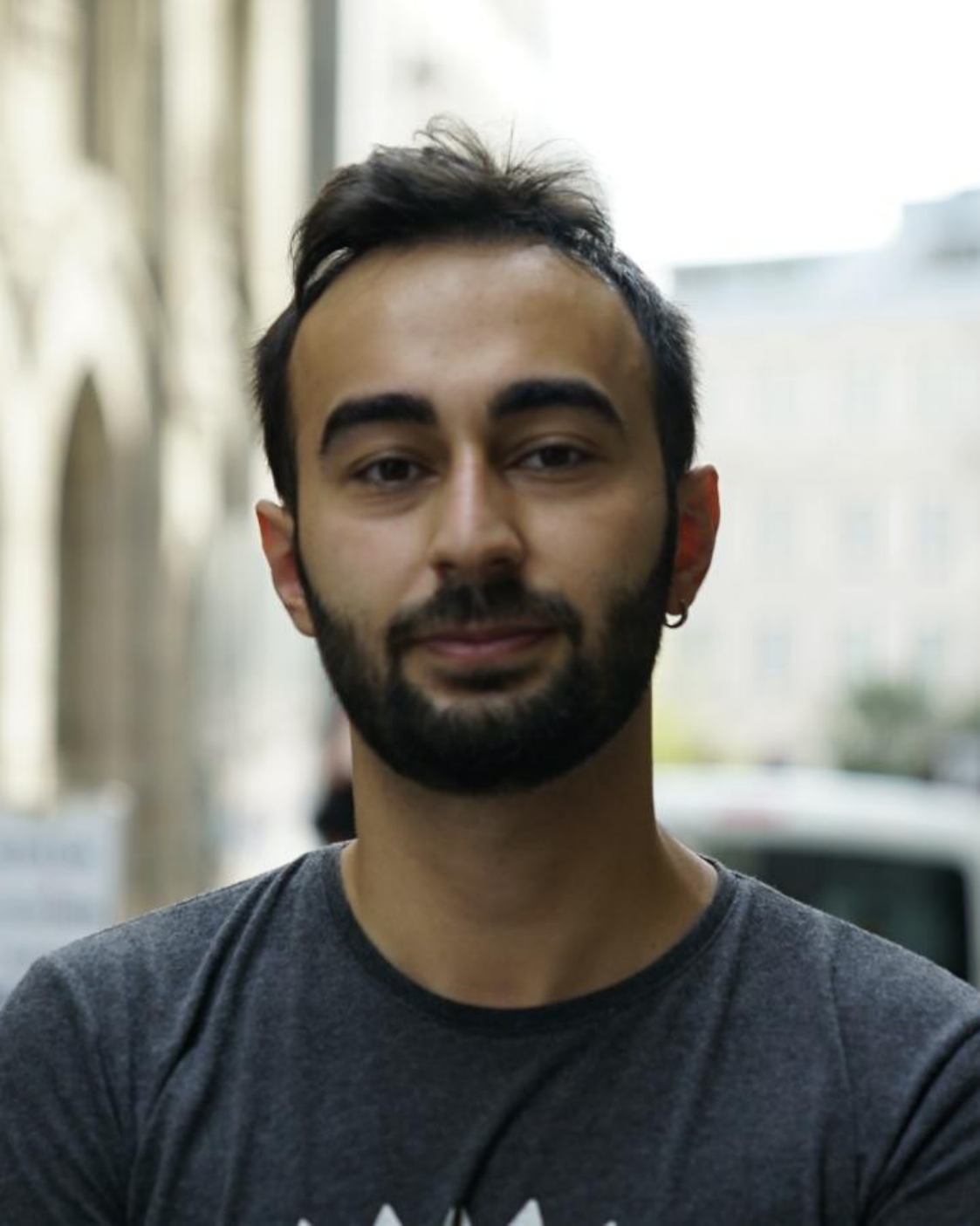}}]{EMIR ASLANDOGAN}
received his B.Sc. degree in Electronics and Communication Engineering from Istanbul Technical University in 2020. He completed his M.Sc. in the Satellite Communications and Remote Sensing program at Istanbul Technical University. Currently, he is pursuing his Ph.D. in the communication program at Yildiz Technical University, where he also serves as a researcher. His research interests include wireless communications, delay-Doppler domain communications, reconfigurable intelligent surfaces, and fluid antennas.
\end{IEEEbiography}
\begin{IEEEbiography}[{\includegraphics[width=1in,height=2in,clip,keepaspectratio]{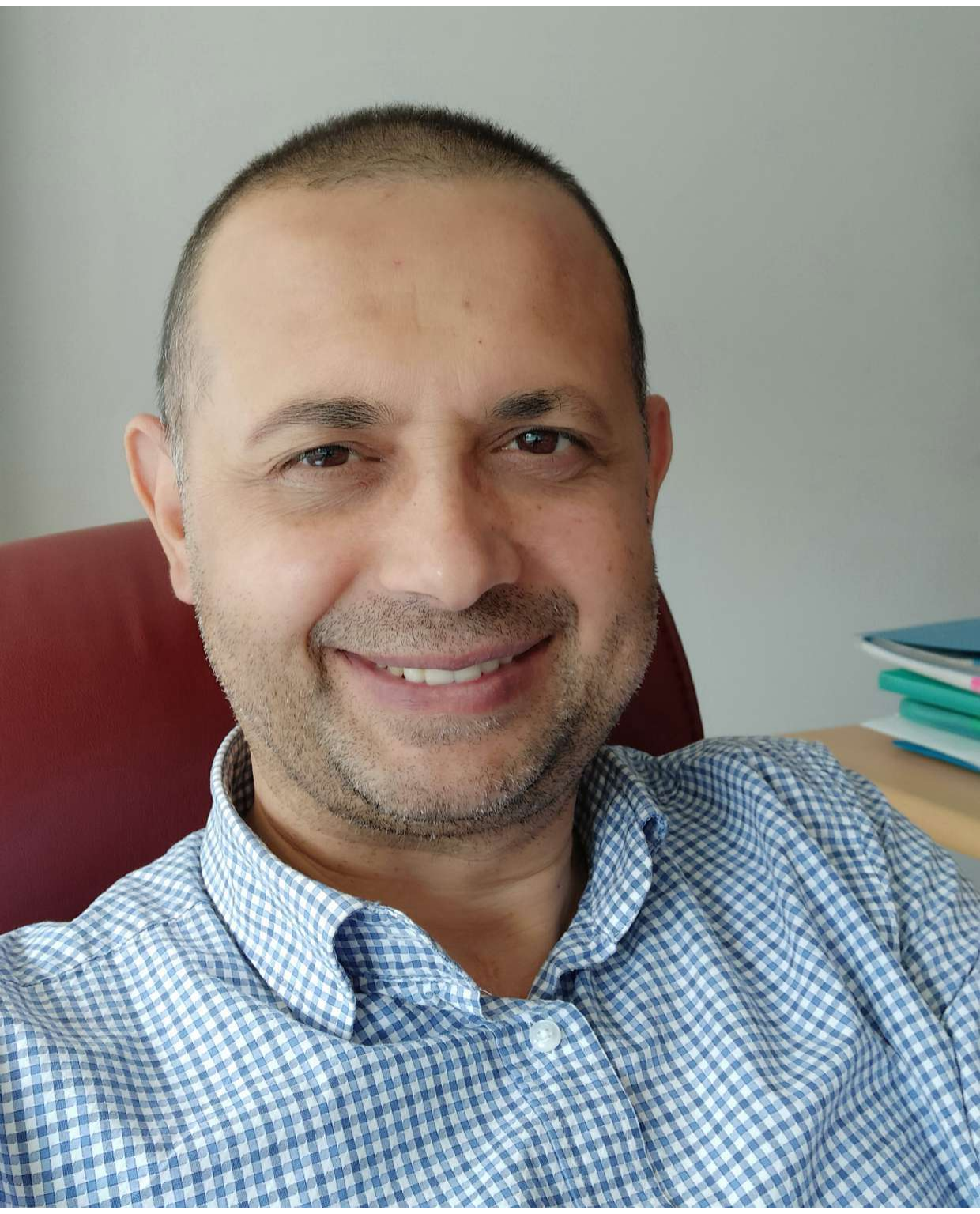}}]{HACI ILHAN} (Senior Member, IEEE) received the B.Sc. degree in electronics and information engineering from Yildiz Technical University, Istanbul, Turkey, and the M.Sc. and Ph.D. degrees in electronics and information engineering from Istanbul Technical University, Istanbul. He was working as a Research and Teaching Assistant with the Department of Communication, Istanbul Technical University, from 2001 to 2011. Since 2011, he has been working as an Associate Professor with the Department of Electronics and Communication Engineering, Yildiz Technical University. His diverse research interests include communications theory and signal processing, with a concentration on wireless technologies, cooperative networking, MIMO communication methods, space-time coding, next-generation communication, and MAC protocols for VANETs are some of the specific study fields.
\end{IEEEbiography}
\vfill\null
\end{document}